\title{Starting with data: advancing spatial data science by building and sharing high-quality datasets}
\author{
        Yingjie Hu \\
                GeoAI Lab, Department of Geography, University at Buffalo,  USA       
}
\date{}

\documentclass[11pt]{article}

\usepackage[round]{natbib}
\usepackage{graphicx}
\usepackage[margin=1in]{geometry}
\usepackage{setspace}
\begin{document}
\maketitle




\vspace*{-0.5cm}
\textit{Spatial data science} has emerged in recent years as an interdisciplinary field. Since \textit{spatial data science} seems to largely overlap with the existing \textit{geographic information science}, it is worth to  examine the difference between these two  names first. Both share the term ``science", and in the context of geographic research, the meanings of ``spatial" and ``geographic" are very close (although the term ``spatial" is more general and has been used in many other disciplines, such as psychology and mathematics). Thus, the main difference between these two names should be the use of the term ``data" instead of ``information".

If we trace the creation of the name ``geographic information science" to Michael Goodchild's 1992 paper \cite[]{goodchild1992geographical}, it seems that one reason he used ``geographic information science" instead of ``geographic data science" was to emphasize that this is a science about geographic information systems. Also, using ``geographic information science" allows the reuse of the same acronym of ``GIS" that was already accepted by many people at that time. Interestingly, if we read the proposed content of geographic information science in section 3 of that paper, it was largely about ``data" rather than ``information", such as data collection, data capture, data modeling, and data structure. In fact, the term ``data" was used 130 times in this 1992 landmark paper of Michael, while the term ``information" was used only 54 times, less than half of the frequency of ``data". In addition, a major GIS conference at that time was \textit{Spatial Data Handling}, and the name of the conference was not \textit{Spatial Information Handling}.  Nevertheless, it was the term ``information" that  eventually made into the name that defines an important field. Another reason could be related to the DIKW pyramid, i.e., Data, Information, Knowledge, and Wisdom. Since we are usually more concerned about getting information (and knowledge) from data rather than simply having the data themselves, we might prefer to use the term ``information". Thus, ``geographic information system" probably sounds more intelligent than ``geographic data system".

Then, why do we start to treat the term ``data" more favorably in recent years, and even use it in the name ``spatial data science"? On the one hand, it is likely that this has something to do with the recently popular field ``data science". While it is still impossible to reach a consensus on the definition of ``data science" (and there are still debates on the legitimacy of such a field), we can't ignore its high popularity among both academia and industry and its significant impact on the society. On the other hand, the preference to the term ``data" should also be attributed to the indeed increased importance of \textit{data} in a wide range of domains. First, the unprecedented volume,  variety, and velocity of today's big data greatly facilitate scientific discovery and give birth to the fourth paradigm of science. Second, data no longer only serve as the resources waiting to be mined by tools but are being used for developing these tools. It has long been recognized in the machine learning community that ``Garbage In, Garbage Out", and high-quality  datasets, such as ImageNet, have become critical enablers for new methods.  

Given the importance of data, it seems that a reasonable starting point for advancing spatial data science is to build and share high-quality datasets. Such a practice has several values for fostering the development of this new field. First, it enables the innovative re-use of existing datasets and contributes to the discovery of new knowledge. A dataset originally collected for answering one research question may be re-used and re-examined from other perspectives, and a synthesis of multiple existing and publicly-shared datasets can produce new knowledge that cannot be obtained by any single dataset before. Second, it reduces duplicated efforts and increases the efficiency of  research and education for the  community. Many high-quality datasets require a considerable amount of time, labor, and financial resources to create. This includes not only cleaning and organizing datasets  but also labeling data records for training supervised models. In addition to their obvious research value, these datasets can also facilitate the education of spatial data science by providing students with data resources that they can explore and test. Finally, high-quality and publicly-shared datasets enhance the reproducibility and replicability of published research, and are critical for the long-term prosperity of a scientific discipline.

What should we do in order to build and share high-quality spatial datasets? We may need to start by defining what \textit{high-quality} means and developing a formal guidance on creating high-quality datasets. While there already exist various standards for geospatial data from FGDC and ISO, such a guidance may be designed in a more inclusive manner to encourage data sharing from more researchers. One example that we can learn from is the 5-star Rating System for sharing Linked Open Data \cite[]{linkeddata5star}, which does not require a dataset to satisfy the highest standard before it is shared but encourages people to simply make data available first on the Web under an open license.  Spatial datasets come with their unique properties, such as their locations, timestamps, and spatial resolutions, whose inclusion should be discussed in the formal guidance. Accordingly, it is necessary to define ``high-quality" for not only \textit{data} but also \textit{metadata}. With a formal guidance for data sharing, a further step we could do is to develop automatic methods that can help existing datasets reach a higher level of quality. In a previous work, we  developed an approach that leverages the Labeled Latent Dirichlet Allocation (LLDA) model to extract standard topics from the existing descriptions to enrich the datasets \cite[]{hu2015metadata}. Similar approaches can be developed to improve the quality of existing datasets. Finally, spatial data infrastructures (SDIs) are likely to play even more important roles in the context of spatial data science. We will need new methods and technologies that can further improve the search and discovery of the high-quality datasets shared in SDIs.

\citet{luc2019} defines \textit{spatial data science} as ``a subset of generic `data science' that focuses on the special characteristics of spatial data". \citet{singleton2019geographic} proposes \textit{geographic data science} which  ``combines the long-standing tradition and epistemologies of Geographic Information Science and Geography with many of the recent advances that have given Data Science its relevance in an emerging `datafied' world". No matter which name we use, it seems that the increased importance of \textit{data} has already been recognized by the community. Building and sharing high-quality datasets could be a starting point for us to advance this emerging field.

\small
\bibliographystyle{plainnat}
\bibliography{GIR18_bibliography}

\end{document}